\documentclass[a4paper]{article}

\usepackage{INTERSPEECH2020}

\usepackage[noadjust]{cite} 
\usepackage{caption, subcaption} 

\title{Speaking Speed Control of End-to-End Speech Synthesis\\using Sentence-Level Conditioning}
\name{Jae-Sung Bae, Hanbin Bae, Young-Sun Joo, Junmo Lee, Gyeong-Hoon Lee, Hoon-Young Cho}
\address{Speech AI Lab, NCSOFT, Republic of Korea}
\email{\{jaesungbae, bhb0722, ysjoo555, ljun4121, ghlee0304, hycho\}@ncsoft.com}

\begin{document}

\maketitle
\begin{abstract}
This paper proposes a controllable end-to-end text-to-speech (TTS) system to control the speaking speed (speed-controllable TTS; SCTTS) of synthesized speech with sentence-level speaking-rate value as an additional input. The speaking-rate value, the ratio of the number of input phonemes to the length of input speech, is adopted in the proposed system to control the speaking speed. Furthermore, the proposed SCTTS system can control the speaking speed while retaining other speech attributes, such as the pitch, by adopting the global style token-based style encoder. The proposed SCTTS does not require any additional well-trained model or an external speech database to extract phoneme-level duration information and can be trained in an end-to-end manner. In addition, our listening tests on fast-, normal-, and slow-speed speech showed that the SCTTS can generate more natural speech than other phoneme duration control approaches which increase or decrease duration at the same rate for the entire sentence, especially in the case of slow-speed speech.

\end{abstract}


\section{Introduction}
In the past few years, the naturalness of synthesized speech has been significantly improved owing to the advancement in neural text-to-speech (TTS) methods \cite{tacotron1, tacotron2, deepvoice3, DCTTS}. Speech varies in expressions; however, these models only focus on the generation of narrative-style speech. Therefore, many researches have been recently proposed to control the prosody and speaking speed of the synthesized speech in a TTS system \cite{prosody-tacotron, amazon-fine-graned, neosapience-robust-and-fine-grained, GST, kiast-duration,fastspeech}. This paper focuses on the control of speaking speed that is essential for real scenario because the speaking speed must vary depending on the context or situation.

In \cite{prosody-tacotron, amazon-fine-graned, neosapience-robust-and-fine-grained,GST}, various speech attributes, such as pitch, prosody, and speaking speed are extracted from a reference speech and the TTS system generates a similar style of speech to the reference speech. Because it is difficult to define every speech attribute objectively, the TTS system is trained in an unsupervised manner. However, because of this unsupervised learning mechanism, to successfully train and control speech attributes using these systems, the training database should contain a wide variety of styles of speech, and the amount of data in the database should be sufficient. Furthermore, it is very difficult to perfectly disentangle each speech attribute from speech (i.e., when we attempt to control the speaking speed, other speech attributes, such as pitch, may also be modified).

In \cite{kiast-duration, fastspeech}, neural TTS systems that control the phoneme-level speech duration have been proposed. Phoneme duration is additionally inputted to the TTS system \cite{kiast-duration}, or the hidden states of the phoneme sequence are expanded, corresponding to the phoneme duration \cite{fastspeech}. These systems, in the inference step, control the speaking speed by modifying the phoneme duration predicted by a duration predictor. However, these types of techniques have two constraints. First, they require an external well-trained automatic speech recognition (ASR) or TTS model to extract the ground-truth phoneme duration from the speech signal; however, to obtain these well-trained models, a database containing a large amount of extra speech data and domain knowledge are also required. Second, it is difficult to control the speaking speed naturally. These systems control the speaking speed by increasing or decreasing the predicted duration of each phoneme at the same rate for an entire sentence. However, when people are asked to speak quickly or slowly, their speaking speed is not consistent for the entire sentence; they control the speed by speaking specific phones quickly and others slowly \cite{SR-intro2}. In \cite{SR-intro1}, the authors reported that the ratio of the duration of the vowels to that of the consonants increased as the speaking speed changed from normal to slow.

To overcome these constraints, we propose a speed-controllable TTS (SCTTS) system that adopts a sentence-level speaking rate (SR) as the input. The SR value is simply calculated as the ratio of the number of input phonemes to the length of input speech for each sentence. During training, by providing the SR value as the input, the proposed system learns to predict different alignments with the same text depending on desired speaking speeds. In the inference step, an average SR value obtained from a training database is used to generate a speech with normal speed, and the increased (decreased) SR value is used to generate fast (slow) speech. Furthermore, we improved the robustness of the speaking speed control of the proposed system by adopting a global style token (GST) \cite{GST}. In our pilot study, we found that the speed-control of a TTS system is affected by other speech attributes when the speech database contains various styles of speech samples (e.g., in addition to the speaking speed, other attributes such as pitch are changed). However, the GST-adopted SCTTS (SCTTS-GST) system can control the speaking speed while minimizing the changes in other speech attributes by disentangling it from multiple speech attributes included in the expressive speech. In the inference step, by providing a normal-style reference speech, together with the SR value, the SCTT-GST can control the speaking speed while retaining other speech attributes of a normal-style speech.

The key strengths of the proposed SCTTS system are as follows. First, the SCTTS system does not require extra labor-consuming well-trained models to extract the ground-truth phoneme duration and can be trained in an end-to-end manner. Second, the speed-controlled speech generated by the SCTTS system is more natural than that generated using other approaches of increasing or decreasing the phoneme duration at the same rate for the entire sentence. Third, the proposed SCTTS-GST system can control the speaking speed while retaining other speech attributes of a normal-style speech even for an expressive dataset.

\section{Baseline Model}
In this section, we present some background knowledge to understand our proposed method effectively.
\subsection{End-to-End TTS Framework}
As an end-to-end TTS framework, we used the deep convolutional TTS (DCTTS) system \cite{DCTTS} with some modifications. Because the DCTTS system is fully convolutional, it has advantages of fast training speed and stable convergence. First, the text-to-mel spectrogram (T2M) network generates the coarse mel spectrogram, which is a down-sampled mel spectrogram in the time axis, from the input text. In the T2M network, the input phoneme (or character) embeddings and input acoustic features represented as the mel spectrogram are encoded by a text and an audio encoder, respectively. The attention module aligns the text and audio encoding, and the audio decoder predicts the coarse-mel spectrogram from the attention module output. Second, the post-processing network (PostNet) transforms the predicted coarse mel spectrogram into a mel spectrogram that is up-sampled in the time axis. As we synthesize speech using a neural vocoder for high-quality speech, the spectrogram super-resolution network (SSRN), which converts a coarse mel spectrogram into a spectrogram in \cite{DCTTS}, is replaced by the PostNet. Finally, a neural vocoder synthesizes the speech waveform by inputting the predicted mel spectrogram.

Unlike the original DCTTS system, which trains T2M network and SSRN independently, we jointly train the T2M network and the PostNet in an end-to-end manner using the following loss function:

\begin{equation}
\mathcal{L} = \alpha \mathcal{L}_{spec}(\bm{\hat{c}}| \bm{c}) + \mathcal{L}_{spec}(\bm{\hat{m}}| \bm{m}),
\label{eq5}
\end{equation}
where the $\bm{\hat{c}}$ and $\bm{c}$ denote the predicted and ground-truth coarse mel spectrograms, and $\bm{\hat{m}}$ and $\bm{m}$ are the predicted and ground-truth mel spectrograms. $\alpha$ is the weight of the coarse mel spectrogram reconstruction loss. The spectrogram reconstruction loss function, $\mathcal{L}_{spec}$, is defined as the summation of L1 loss and binary divergence $D_{bin}$ as in \cite{DCTTS}. 

\subsection{GST-based Style Encoder}
The GST-based style encoder extracts the speaking style from the reference speech as a style embedding. First, the input reference speech signal is encoded by the reference encoder that was proposed in \cite{prosody-tacotron}. Then, the attention module generates the weights between the reference embedding and the GSTs, which represent their similarities. Finally, the weighted summation of GSTs forms the style embedding. In the training stage, the GSTs are randomly initialized and learned to contain the speech styles that appear dominant in the training dataset in an unsupervised manner. In the inference stage, the styles of synthesized speech can be controlled by the weights of GSTs. They can either be obtained by inputting the reference speech signal or be given directly.

\section{Proposed Method}
\subsection{Speed-Controllable TTS Model}
\begin{figure}[t]
  \centering
  \includegraphics[width=\linewidth]{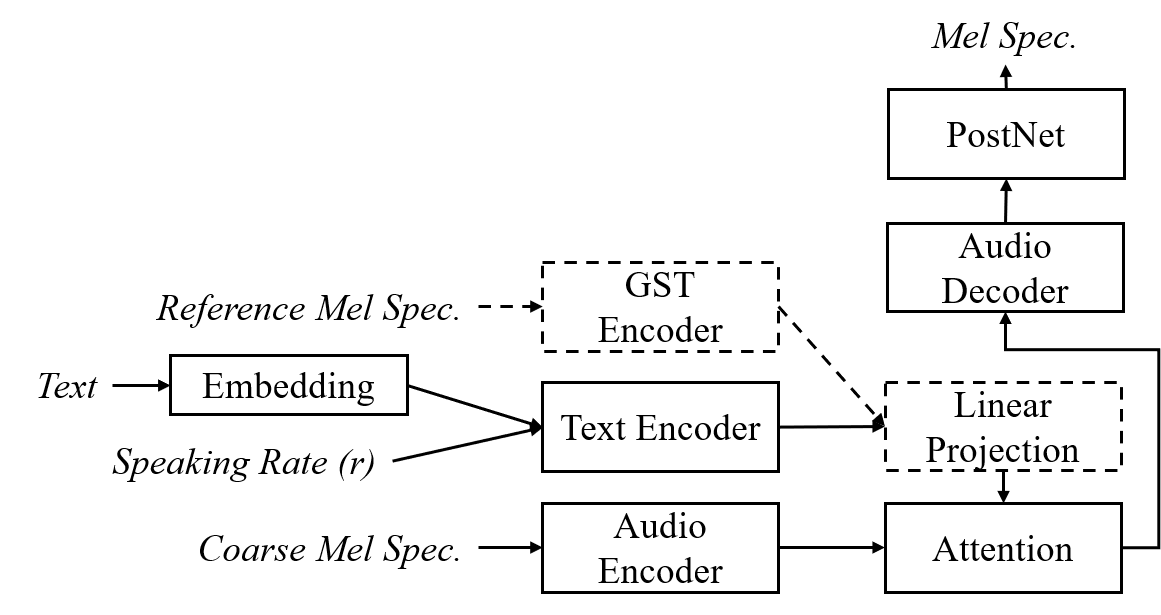}
  \caption{Architecture of our proposed SCTTS system. Dotted-line components are added for the SCTTS-GST.
  }
  \label{fig:model}
\end{figure}

To control the speaking speed of the synthesized speech, we proposed the SCTTS system, which applies the sentence-level SR value to the end-to-end TTS system as an additional input. The architecture of the SCTTS system is detailed in Figure \ref{fig:model}. 

In the training stage, the text sequence, coarse mel spectrogram, and computed SR value are given as inputs. The SR value is replicated to a text-embedding sequence length, concatenated with it, and then given as an input to the text encoder. In the inference stage, text sequences and the desired SR value are given, and the mel spectrogram with corresponding speaking speed is predicted.

\subsection{Sentence-level SR Calculation}
There are several methods of computing the SR value \cite{speech-rate-MR, speech-rate-IMD, speech-rate-recent}. We used the inverse mean duration (IMD) \cite{speech-rate-IMD} as the SR measurement. For each sentence, when the mel spectrogram with a length of $T$ and the text sequences with the number of $P$ are given, the SR value can be defined as follows:
\begin{equation}
r = \lambda \cfrac{P}{T},
  \label{eq1}
\end{equation}
where $\lambda$ is a SR scaling factor. Please note that the silence regions are removed. 

\subsection{Disentanglement of Speaking Speed and Other Speech Attributes}
\begin{figure}[t]
  \centering
  \includegraphics[width=0.8\linewidth]{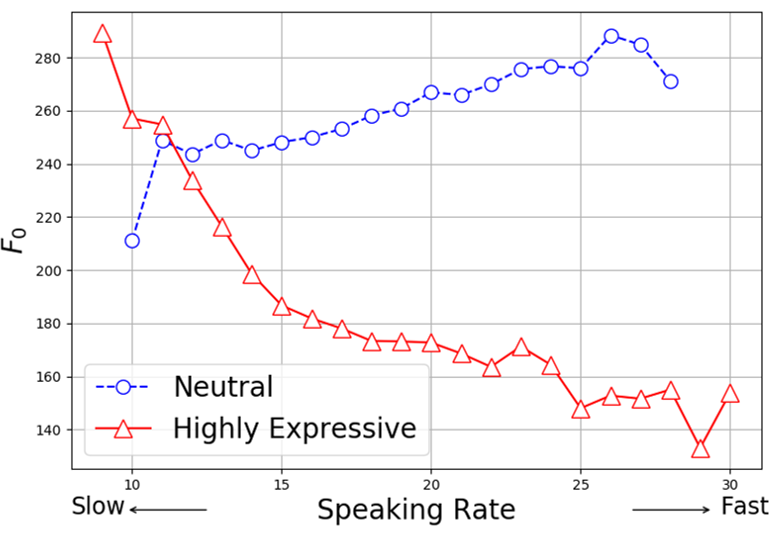}
  \caption{$F_0$ vs. SR value of (red triangle) neutral-speaker and (blue circle) highly expressive-speaker training dataset.}
  \label{fig:train_anal}
\end{figure}

We observe that multiple speech attributes are correlated when the speaking style of the speech dataset has significant variations. Figure \ref{fig:train_anal} depicts the relationship between fundamental frequency ($F_0$) and SR of our neutral and highly expressive speech datasets; the speech dataset is detailed in section 4.1. In Figure \ref{fig:train_anal}, in our neutral dataset, the $F_0$ remains almost unchanged as the speaking speed changes, however, in our expressive dataset, it changes significantly. This type of correlation among speech attributes ($F_0$ and SR in Figure \ref{fig:train_anal}) affects speed control in the SCTTS system.

To disentangle and control the speaking speed from other speech attributes, we adopted a GST-based style encoder \cite{GST}. Because a GST-based style encoder learns style features in an unsupervised manner, in the SCTTS-GST system, it is expected to learn style features other than the speaking speed, which is provided as the SR value. The additional components of the SCTTS-GST system are indicated by dotted lines in Figure \ref{fig:model}. The style embedding, which is the output of the GST-based style encoder, is concatenated with the text encoder output and projected through a linear layer. The other processes and loss function are the same as those of the basic SCTTS system.

\section{Experiments and Results}
We experimentally evaluated the following three aspects of the SCTTS system: 1) controlling the speaking speed, 2) disentangling the speaking speed from other speech attributes using GSTs, and 3) achieving naturalness of the synthesized speech with various speeds. Audio samples can be found online\footnote{
https://nc-ai.github.io/speech/publications/speed-controllable-tts}.

\subsection{Datasets}
For the experiments, we used two Korean speech datasets; one neutral and one highly expressive. The neutral speech dataset, KSS Dataset \cite{KSSdataset}, was recorded by a professional female voice actress. The total amount of speech is about 8.7 h. The expressive speech dataset, an internal dataset, was recorded by a professional male voice actor. This dataset consists of speech samples with various expressions (neutral, excited, shouting, sigh, etc). The total amount of speech is 11 h. Both datasets were recorded in a studio environment and the frequency sampling rate is 22050 Hz. For testing, $1\%$ of each dataset was randomly selected. We used phoneme as the text input and the mel spectrogram with 80 bins computed with an FFT size of 1024, a hop size of 256, and a window size of 1024, as the acoustic feature.

\subsection{Model Setup}
The architectures of the text encoder, audio encoder, and audio decoder were the same as in \cite{DCTTS}. The architecture of PostNet was the same as that of the SSRN in \cite{DCTTS}, but the output frequency bin was set to 80 because the PostNet predicts mel spectrogram. The SR scaling factor, $\lambda$, was set as 100. The loss weight, $\alpha$, was set as 1 before 50 k training steps; it decreased linearly and became 0 at the 200 k step. We did not use the guided attention loss in \cite{DCTTS} because the system was trained stably and fast even without the guided attention loss. The Adam optimizer \cite{ADAM} was used to optimize the network with an initial learning rate of 0.001. For the GST-based style encoder, the same network architecture with ten style tokens and a multi-head attention module with four heads as in \cite{GST} were used. For a neural vocoder, WaveGlow \cite{waveglow} was used, and it was trained using a database containing approximately 50 h of speech recorded by four speakers.

\subsection{Speaking Speed Control}

\begin{figure}[t]
    \centering
        \centering
        \includegraphics[width=\linewidth]{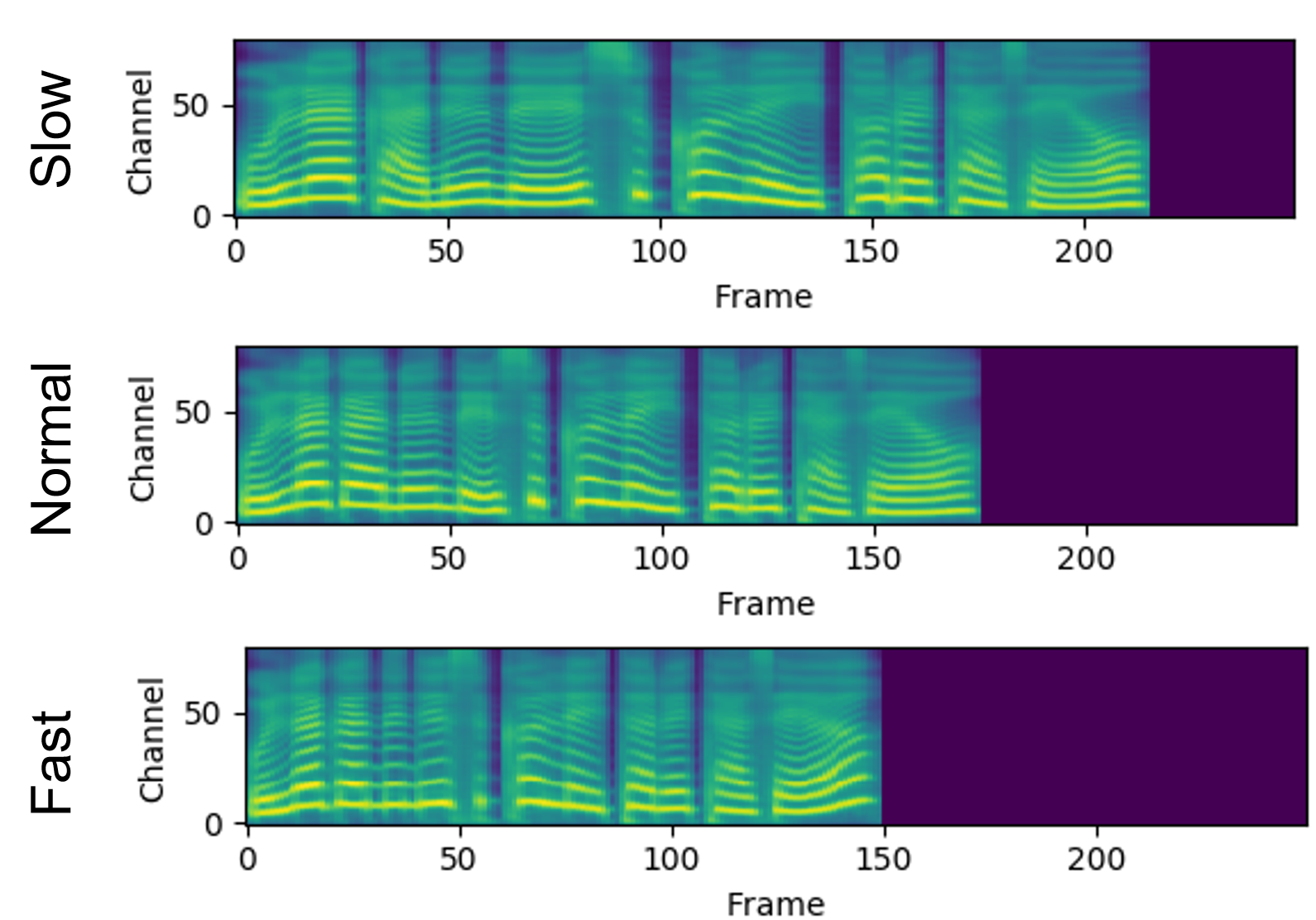}
        \caption{Mel spectrograms of the synthesized speech with different speaking speeds. These were generated from the SCTTS system trained with the neutral speech dataset.}
  \label{fig:result1-a}
\end{figure}

Figure \ref{fig:result1-a} shows the synthesized speech samples with various speaking speeds generated by the SCTTS system trained with the neutral speech dataset. We controlled the speed of the speech by adjusting the SR value. The average SR value representing the normal speed was obtained from the training set. As depicted in Figure \ref{fig:result1-a}, the SCTTS system naturally controlled the speaking speed while almost maintaining other speech attributes such as pitch, without leading to any distortions such as trembling or breaks in the mel spectrogram.

\subsection{Disentanglement of Speaking Speed and Other Speech Attributes}
\begin{figure*}[t]
    \centering
    \begin{subfigure}{0.34\linewidth}
        \centering
        \includegraphics[width=\linewidth]{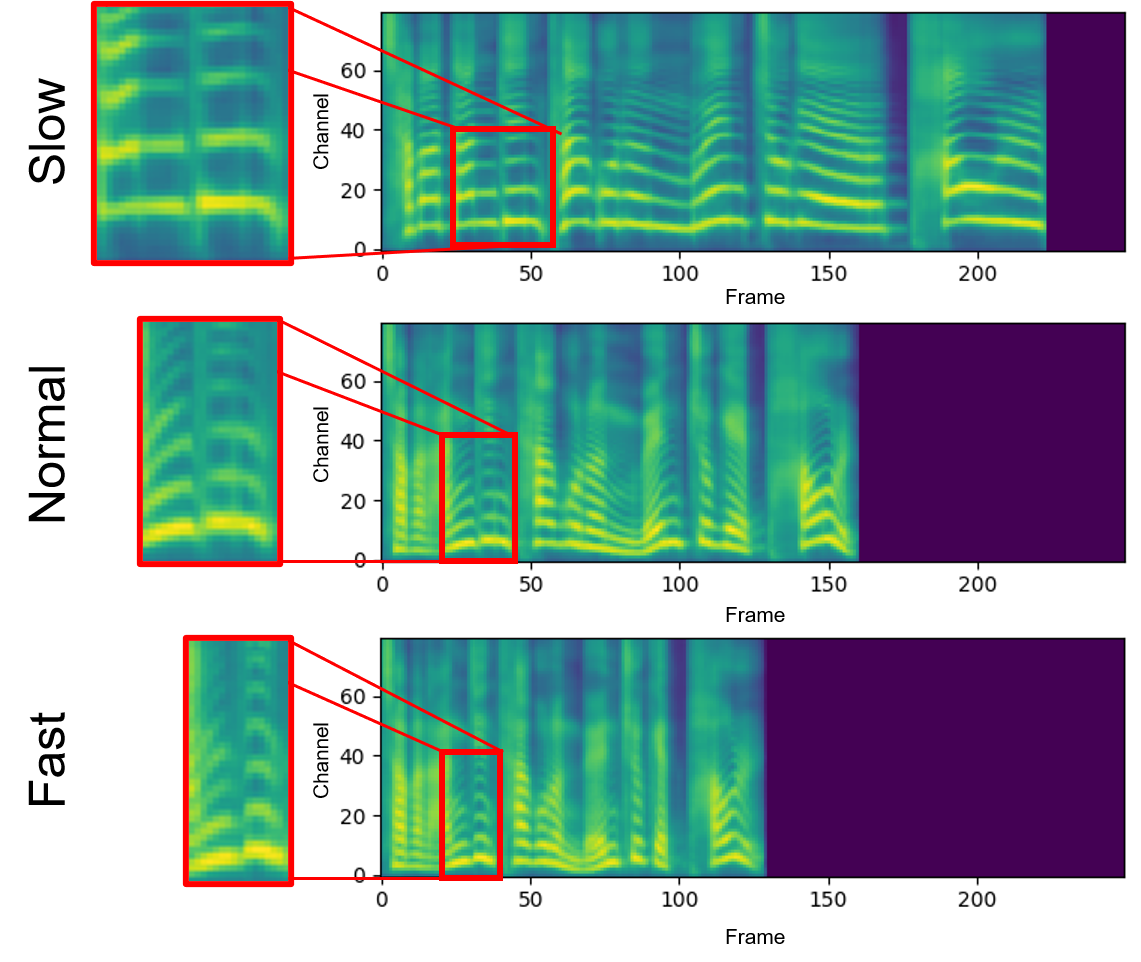}
        \caption{SCTTS}
        \label{fig:result1-b}
    \end{subfigure}
    \begin{subfigure}{0.32\linewidth}
        \centering
        \includegraphics[width=\linewidth]{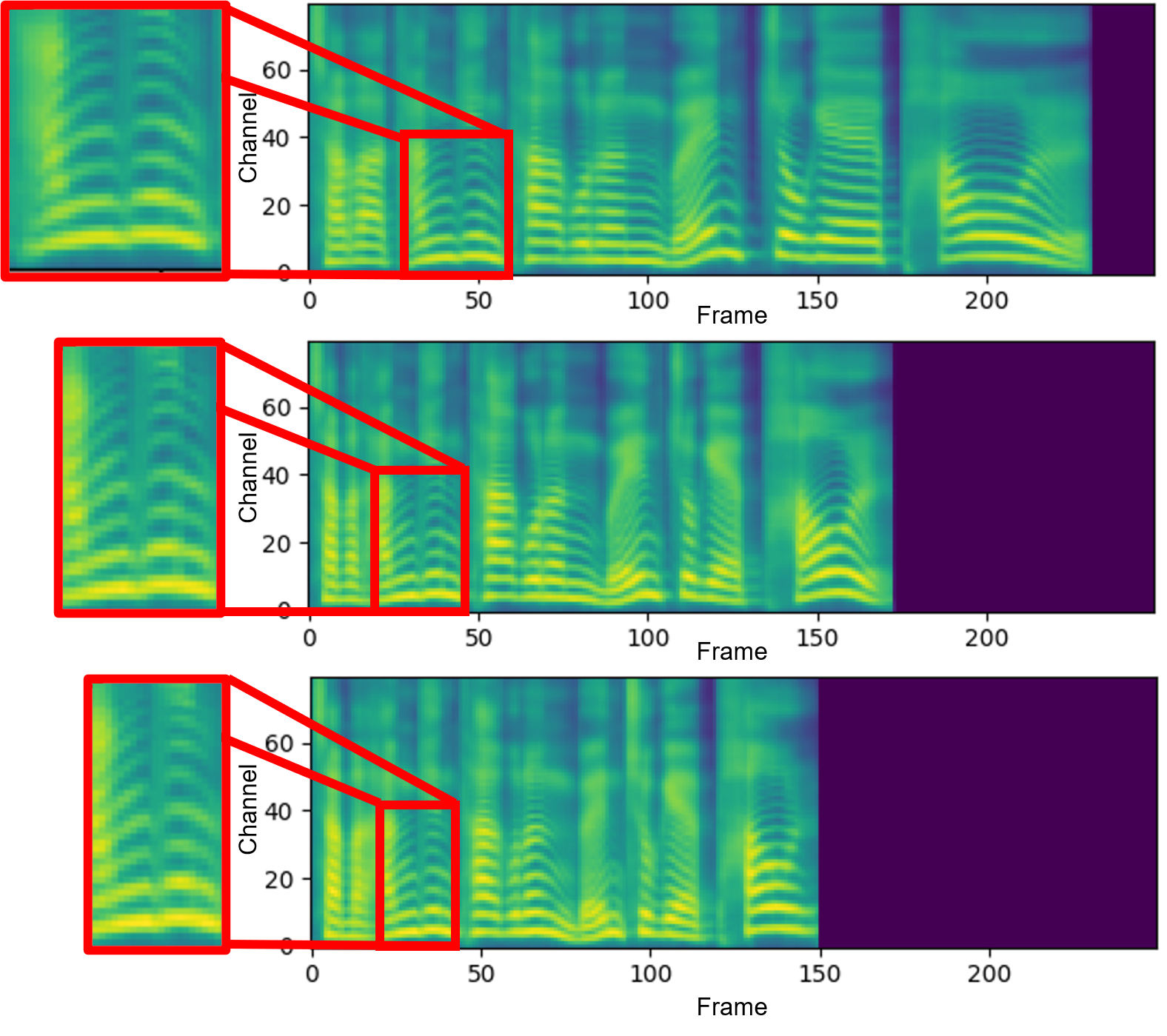}
        \caption{SCTTS-GST (N)}
        \label{fig:result1-c}
    \end{subfigure}
    \begin{subfigure}{0.33\linewidth}
        \centering
        \includegraphics[width=\linewidth]{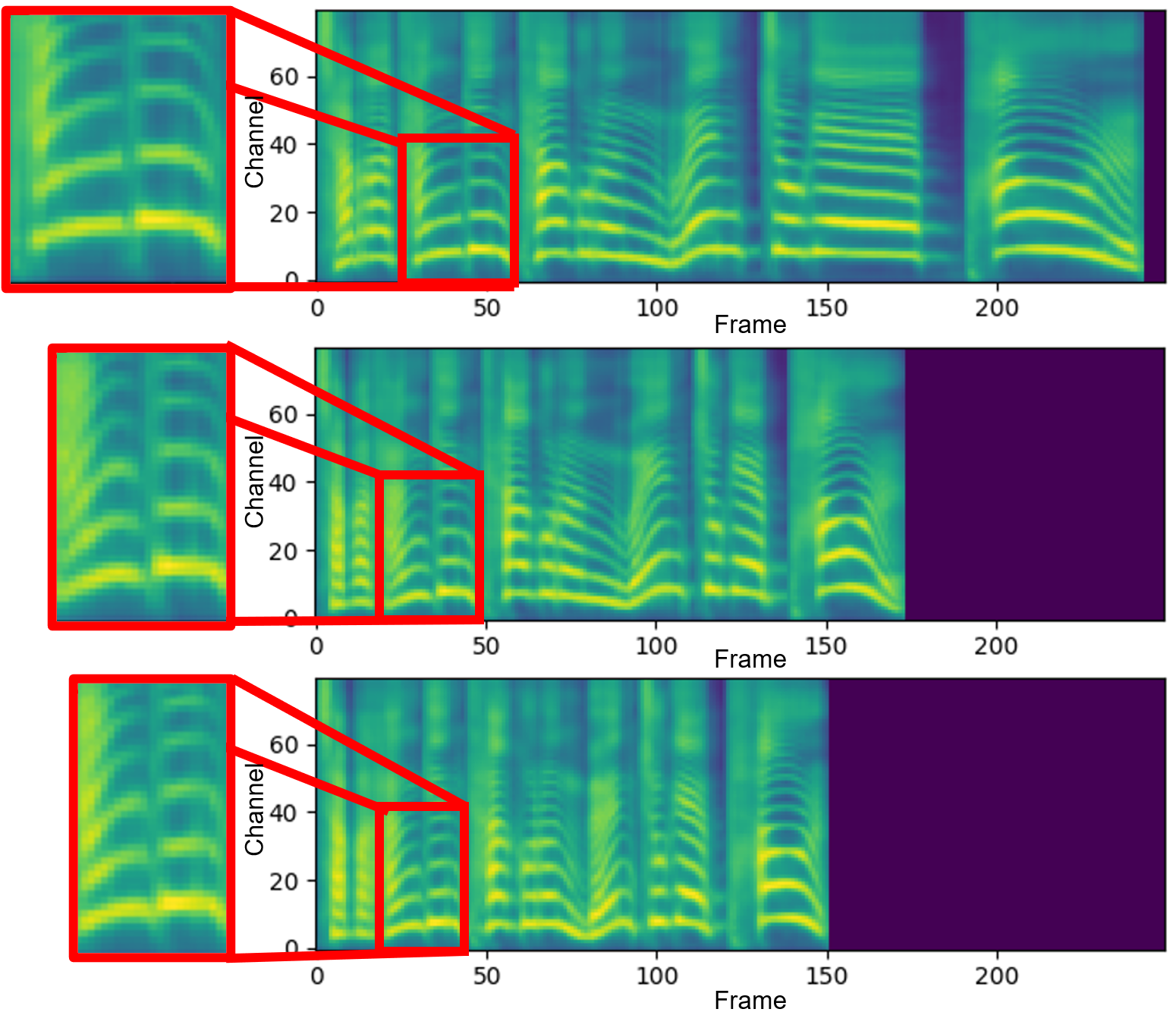}
        \caption{SCTTS-GST (H)}
        \label{fig:result1-d}
    \end{subfigure}
    \caption{Mel spectrograms of the synthesized speech from the SCTTS and SCTTS-GST systems with different speaking speeds. ``N'' and ``H'' in the SCTTS-GST represent the normal- and high-$F_0$ reference speech cases, respectively. In (a), $F_0$ increased as the SR value decreased, depending on the speaker characteristics; however, in (b) and (c), by applying the GST technique, we successfully disentangled a person's speaking speed from $F_0$.
    }
    \label{fig:result1}
\end{figure*}

\begin{figure}[t]
    \centering
        \centering
        \includegraphics[width=0.85\linewidth]{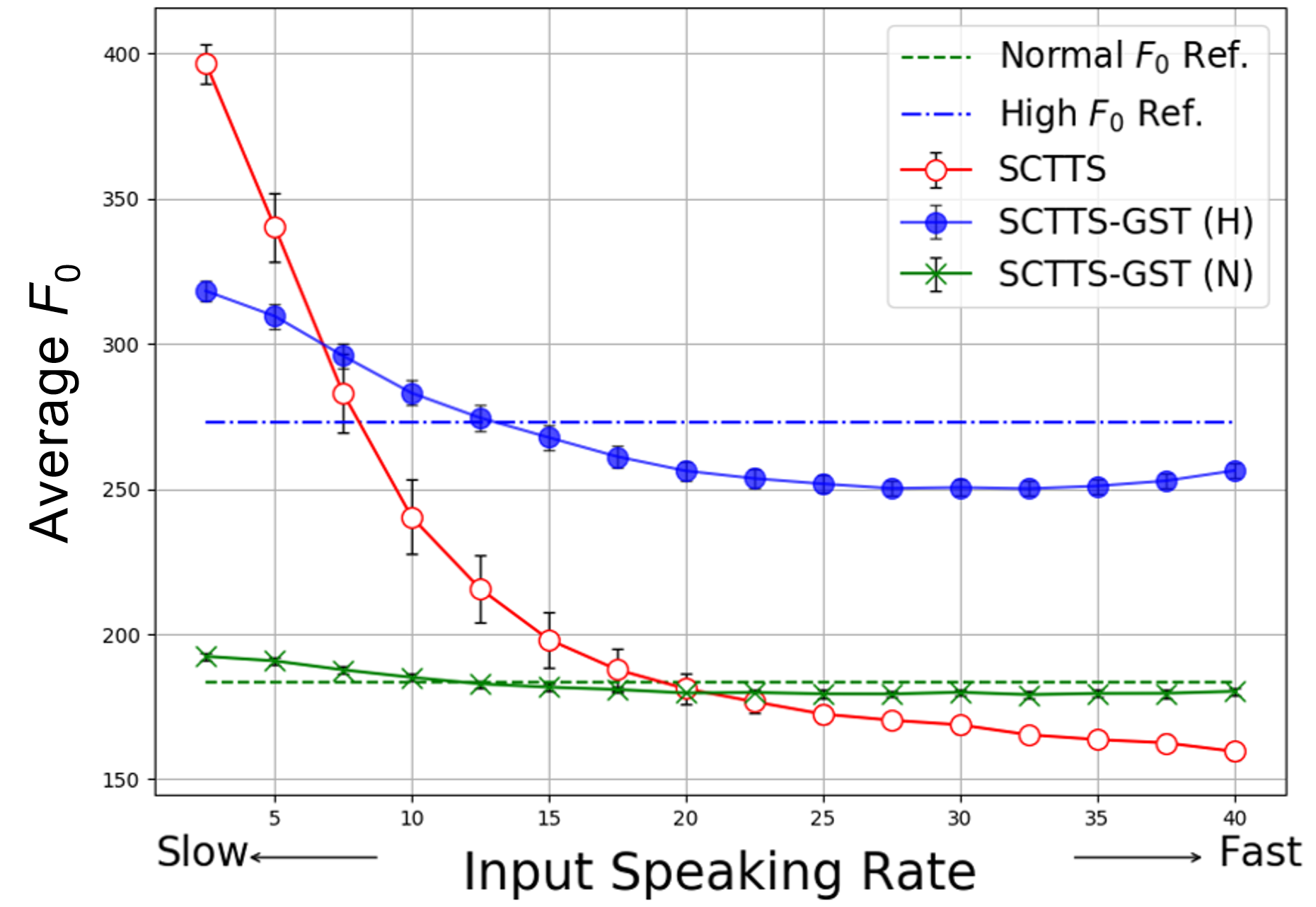}
        \caption{Average $F_0$ vs. input SR value for the synthesized speech of the test set with 95\% confidence interval (CI).}
  \label{fig:result_anal}
\end{figure}

In this sub-section, we evaluated the proposed system in the aspect of disentanglement between the speaking speed and other speech attributes. Figure \ref{fig:result1} shows examples of speech samples with various speeds synthesized by the SCTTS and the SCTTS-GST systems. Both the systems were trained using the expressive speech dataset. For the SCTTS-GST system, we used two different reference speech, one with normal style (non-expressive; average $F_0$) and the other with excited style (high $F_0$). As illustrated in Figure \ref{fig:result1}, in the mel spectrograms of the synthesized speech from the SCTTS-GST system with normal and excited reference speech, $F_0$ remains unchanged as the speaking speed changes, whereas in the SCTTS system, $F_0$ changes depending on the speaking speed.

Figure \ref{fig:result_anal} illustrates the average $F_0$ of the synthesized speech according to the SR values. For the test set, we synthesized speech samples for the same text by changing the SR value. As depicted in Figure \ref{fig:result_anal}, in the SCTTS, the $F_0$ was changed significantly according to the SR value. However, in the SCTTS-GST, the $F_0$ was nearly maintained regardless of the SR value (SCTTS-GST (N)) or changed much less (SCTTS-GST (H)) than the SCTTS. Because the speaking speed and high-$F_0$ are more strongly correlated in the training dataset (Figure \ref{fig:train_anal}).

\subsection{Naturalness of the Speed-Controlled Speech}
We evaluated the naturalness of the proposed system and compared it with phoneme-level speech duration control TTS systems. We compared the SCTTS and the SCTTS-GST systems with the phonemic-level duration controllable TTS (PDC-TTS) system \cite{kiast-duration} and the FastSpeech \cite{fastspeech}. Please note that the same WaveGlow neural vocoder \cite{waveglow} was used in all the systems for high-quality speech generation.

\subsubsection{Comparison Model Setup}
\textbf{PDC-TTS: } For a fair comparison, we implemented the TTS part of the PDC-TTS by replacing it with DCTTS instead of Tacotron \cite{tacotron2}. Accordingly, we concatenated the duration embedding, defined in \cite{kiast-duration}, to the text encoder output, and fitted the output dimension through a linear projection. Because the phoneme duration model (i.e., duration predictor), which is essential in the speech-synthesis phase, is not mentioned in \cite{kiast-duration}, we adopted the phoneme duration model in \cite{deepvoice1}, except that the $F_0$ feature is removed from the input and output of the model. To train the duration model, the ground-truth phoneme duration was extracted from the speech database using a Kaldi-based ASR model \cite{Kaldi}; the ASR model was pre-trained using an external ASR speech database. The trained phoneme duration model has a mean absolute error of 21.42 ms for the test set.
\\

\noindent
\textbf{FastSpeech: } We used open source implementation in the ESPnet framework \cite{espnet} for training transformer TTS \cite{transformer-TTS} and FastSpeech. The pre-trained transformer-based TTS model was used to extract the ground-truth phoneme duration information.

\subsubsection{Subjective Evaluation}
We conducted MOS tests on the naturalness of the fast-, normal-, and slow-speed speech samples synthesized by each system. The length of the fast- and slow-speed speech were set as 70\% and 150\% of the length of the normal-speed speech, respectively. Each system was trained with the expressive dataset. For each system, fifteen speech samples were generated. A total of 20 native Koreans participated. They were asked to score the naturalness of the synthesized-speech samples from 1 to 5. The naturalness of the speech synthesized by the DCTTS system is also compared for the normal-speed speech.

\begin{figure}[t]
    \centering
        \centering
        \includegraphics[width=\linewidth]{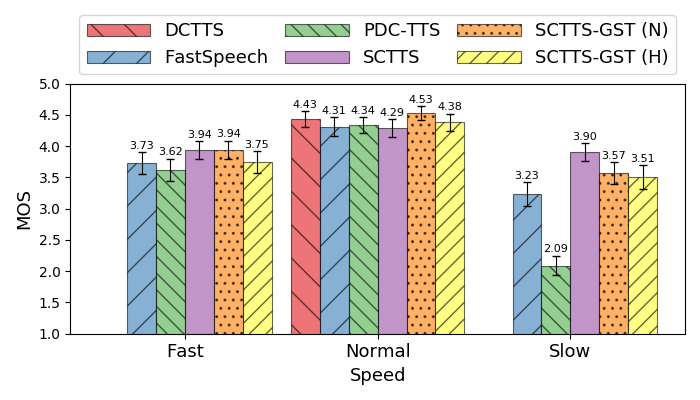}
        \caption{MOS test results on the naturalness with 95\% CI.}
  \label{fig:result_mos}
\end{figure}
The MOS test results are presented in Figure \ref{fig:result_mos}. The performances of the proposed SCTTS and SCTTS-GST systems were comparable to or better than those of other approaches. Furthermore, at a slow speed, the proposed systems outperform other approaches drastically. Because the proposed systems generate slow-speed speech by lengthening the specific phoneme, word, or silence longer than the others, whereas other systems synthesize slow-speed speech by equally lengthening. The unnaturalness of speech generated from other approaches was more noticeable with the slow-speed speech than the fast-speed speech.

\section{Conclusions}
This paper proposed a speed-controllable TTS system that can generate a speech with various speaking speeds without the use of any pre-trained models and can be trained in an end-to-end manner. The proposed SCTTS system can be integrated with the GST and the speaking speed can be disentangled from other speech attributes such as pitch. The proposed SCTTS system outperformed the existing phoneme duration-control TTS systems in MOS listening tests, especially when the speed of the synthesized speech was slow.

\bibliographystyle{IEEEtran}

\bibliography{mybib}

\end{document}